\documentclass[preprint2]{aastex}

\usepackage{natbib, aas_macros}
\shorttitle{The origin of S-stars and a young stellar disk}
\shortauthors{Fujii et al.}

\begin{document}

\title{The origin of S-stars and a young stellar disk: distribution of debris 
stars of a sinking star cluster}


\author{M. Fujii\altaffilmark{1,3}, M. Iwasawa\altaffilmark{2,3},
Y. Funato\altaffilmark{2}, and J. Makino\altaffilmark{3}}

\altaffiltext{1}{Department of Astronomy, Graduate School of Science, The
University of Tokyo, 7-3-1 Hongo, Bunkyo, Tokyo 113-0033;
fujii@cfca.jp}

\altaffiltext{2}{Department of General System Studies, College of Arts and
Sciences, The University of Tokyo 3-8-1 Komaba, Meguro, Tokyo 153-8902; 
iwasawa@cfca.ac.jp, funato@artcompsci.org}

\altaffiltext{3}{Division of Theoretical Astronomy, National
Astronomical Observatory of Japan, 2-21-1 Osawa, Mitaka, Tokyo,
181-8588; makino@cfca.jp}

\begin{abstract}
Within the distance of 1 pc from the Galactic center (GC), more than 100 young 
massive stars have been found. The massive stars at 0.1--1 pc from the GC are 
located in one or two disks, while those within 0.1 pc from the GC, S-stars, have 
an isotropic distribution. How these stars are formed is not well understood, 
especially for S-stars.
Here we propose that a young star cluster with an intermediate-mass black hole 
(IMBH) can form both the disks and S-stars. 
We performed a fully self-consistent $N$-body simulation of a star cluster
near the GC. Stars escaped from the tidally disrupted star cluster were carried to the 
GC due to a 1:1 mean motion resonance with the IMBH formed in the cluster.
In the final phase of the evolution, the eccentricity of the IMBH becomes very high.
In this phase, stars carried by the 1:1 resonance with the IMBH were dropped from 
the resonance and their orbits are randomized by a chaotic Kozai mechanism.
The mass function of these carried stars is extremely top-heavy within $10''$. The 
surface density distribution of young massive stars has a slope of $-1.5$ within 
$10''$ from the GC.
The distribution of stars in the most central region is isotropic.
These characteristics agree well with those of stars observed within $10''$ from 
the GC.

\end{abstract}


\section{Introduction}

More than 100 young massive stars have been found in the Galactic centerr (GC) by 
near-infrared observations 
\citep{1995ApJ...447L..95K,2006ApJ...643.1011P,2006JPhCS..54..279L,
2009ApJ...703.1323D,2010ApJ...708..834B}.
These stars appear to reside in one or two disks at more than $1''$ from the
GC \citep{2006JPhCS..54..279L,2006ApJ...643.1011P}, while B-type stars within 
$1''$ (S-stars) have an isotropic and thermal distribution 
\citep{2003ApJ...596.1015S,2009ApJ...692.1075G}.
Two major scenarios have been proposed for the formation of these stars. 
One is the in-situ formation in an accretion
disk \citep{2003ApJ...590L..33L} and the other is the migration of a star cluster
formed several parsec or more away from the GC \citep{2001ApJ...546L..39G}. 

In the 1990s, only OB and Wolf--Rayet (WR) stars could have been observed in the 
GC. However, recent advance in observation techniques has made it possible to 
identify many late-type stars, and now we can obtain their spacial distribution.
While the OB and WR stars have a power-law distribution with a slope of about 
$-1.5$, the slope for all stars including late-type giants is nearly 0 
\citep{2009ApJ...703.1323D,2009A&A...499..483B,2010ApJ...708..834B}. 
The orbits of S-stars appear to be isotropic and thermal 
\citep{2003ApJ...596.1015S}, but the most recent observation showed that it is more 
eccentric with the distribution of $n(e)\sim e^{2.6}$ \citep{2009ApJ...692.1075G}.
How these distributions have formed is the key to the understanding of the formation 
process of the young stars near the GC.
The mass function of the young stars have also been determined from the observations. 
It is quite top-heavy
\citep{2009ApJ...703.1323D,2009A&A...499..483B,2010ApJ...708..834B}.
Formation scenarios should also explain the mass function.

In-situ formation of S-stars is difficult because of the strong 
tidal field of the central supermassive black hole (SMBH).
Therefore, it is necessary to carry young stars to the current location from
somewhere outside. Recently, several scenarios have been proposed to carry 
stars which were born on a gaseous disk at around 0.1--0.5 pc from the GC and to
randomize their orbits.
The scenarios include migration via the gravitational torques in the stellar 
disk \citep{2007MNRAS.374..515L,2009ApJ...702L...1G}, migration as a star cluster 
core \citep{2006ApJ...650..901B}, randomization of the orbital 
elements of S-stars by an IMBH \citep{2009ApJ...705..361G,2009ApJ...693L..35M} or 
stellar-mass black holes \citep{2009ApJ...702..884P}, 
and the formation of S-stars due to disruptions of binaries 
\citep{2008ApJ...683L.151L,2009ApJ...697L..44M}. 
On the other hand, \cite{2008Sci...321.1060B} suggested the 
direct formation of S-stars in a gaseous disk formed from a giant molecular 
cloud infalling to the GC. However, none of them is well established.

We have shown that star clusters can carry young stars to
the GC, if an IMBH forms in the clusters. In \citet{2009ApJ...695.1421F}, 
we performed a fully self-consistent $N$-body simulation, in which both
the internal dynamics of a star cluster and the interaction between the
cluster and its parent galaxy are handled correctly. The star cluster migrating
into the GC was completely disrupted by the tidal force and the stars escaping from 
the cluster formed a disk structure. Before the disruption, an IMBH formed through the 
runaway collisions of stars in the cluster. We found a new migration mechanism of 
young stars, a 1:1 mean motion resonance with the IMBH. The IMBH carries young stars 
to the GC by the 1:1 resonance after the disruption of the cluster. In this 
simulation, however, 
the spacial resolution around the SMBH was limited to 0.2 pc because of the use of 
a large softening length for the SMBH. Hence, it was impossible to compare the 
distributions of the young stars obtained in the simulation with the observed one.

In this Letter, we report the result of a new simulation performed using our improved 
code which does not need softening for the SMBH. In this simulation, we can follow 
the orbits of stars down to the AU scale, where S-stars reside. We found that the 
distribution of stars obtained from the simulation agrees very well with the 
observations in the following three 
points: the surface density of young massive stars has a slope of $-1.5$,
the young stars have an extremely top-heavy mass function, and 
the orbits of young stars in the inner most region are thermal and isotropic.
A sinking star cluster can explain both a young stellar disk and S-stars at the same 
time.

We describe the method of our $N$-body simulation in Section 2. In
Section 3, we show the results of simulations. Section 4 is for summary.

\section{$N$-body simulation}
We adopted a King model with the non-dimensional
central potential of $W_0=10$ as the Galaxy model. We placed a central 
SMBH with a mass of $3.6\times 10^6 M_{\odot}$ \citep{2005ApJ...628..246E}. Our
galaxy model represents the central region of our Galaxy. 
The total mass of our galaxy model is $5.8\times10^7 M_{\odot}$ in the real scale 
(excluding the SMBH). The number of particles is $6\times 10^6$ and 
the mass of a particle is 9.7 $M_{\odot}$. The half-mass radius is 22 pc and 
the initial core radius before we place the SMBH is 0.8 pc. 
As a model of a star cluster, we also adopted a King model with the
non-dimensional central potential $W_{0}=6$. The number of
particles is 64k. We started a simulation with the initial position of the cluster 
of 12.5 pc from the GC and with an eccentric orbit. 
The initial orbital velocity is 0.67 times the circular velocity. 
The total mass of the cluster is $2.1\times 10^5 M_{\odot}$ and 
the half-mass radius is 0.16 pc. 
Initial mass function (IMF) of stars in the clusters is a Salpeter IMF with lower and 
upper cutoff at 1 and 100 $M_{\odot}$ \citep{1955ApJ...121..161S}. We assigned 
each star a mass randomly chosen from the Salpeter IMF, 
irrespective of its position. In this simulation, we adopted collisions of 
stars in the star cluster, mass-loss from very massive stars, and 
formation of an IMBH in the cluster. The details are described in 
\citep{2009ApJ...695.1421F}.

The calculation was performed using the Bridge code 
\citep{2007PASJ...59.1095F}, which is
a tree-direct hybrid scheme. It can treat dense regions embedded in a
larger-scale system. In this simulation, the internal motion of
star clusters and field stars within 0.28 pc from the central SMBH are
calculated using a direct code, the sixth-order Hermite scheme 
\citep{2008NewA...13..498N}, and other interactions are calculated using a 
treecode \citep{1986Natur.324..446B}. 
The softening length is $3.1\times 10^{-3}$ pc for field stars. We did
not use softening for other interactions. We assume that stars in the distance 
200 times the Schwarzshild radius of the SMBH (0.14 AU) merge to the SMBH.
The step size for the treecode is 15 years. It is short enough to resolve
the orbital motion at the boundary between the tree and direct regions.
Thus, we can follow the orbits of stars down to the AU scale, in which S-stars 
are located. We used the opening angle $\theta = 0.75$ with the center-of-mass 
approximation for the treecode. 
The simulation was performed on the Cray XT4 at National Astronomical
Observatory of Japan. We used 256 cores, and the total CPU time for the calculation 
was about 100 days.

\section{Distribution of young stars in the Galactic center}

The star cluster spiraled into the GC due to the dynamical
friction. At around 1 pc from the GC, the cluster was
completely disrupted by the tidal force of the parent galaxy and the remnant 
stars of the star cluster formed a thick disk. 
Before the disruption, an IMBH formed in the cluster through the runaway
collisions of stars.  A massive star grew via repeated collisions and
finally the mass of the massive star reached about 16000 $M_{\odot}$. 
After the disruption, the IMBH continued to sink to the GC, and a few 
hundred stars which were the members of the cluster also migrated into the GC 
due to the 1:1 mean motion resonance of the IMBH. This evolution is the same as 
that shown in \citet{2009ApJ...695.1421F}, but in the current simulation we can follow
the orbits of the IMBH and stars down to AU scales, while in 
\citet{2009ApJ...695.1421F}
stars could not go below 0.2 pc because of the large softening length.

\begin{figure}[htbp]
\epsscale{0.9}
\plotone{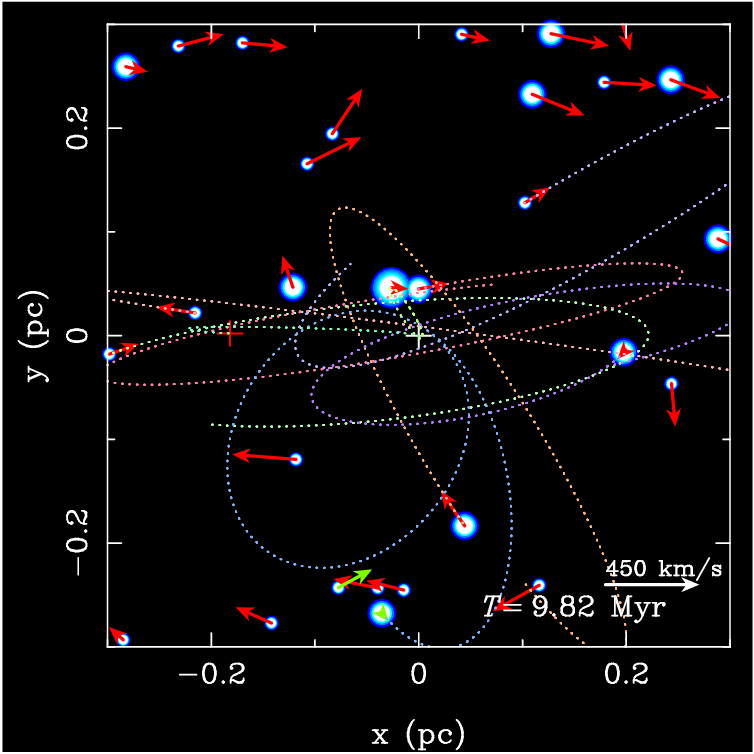} 
 \caption{
   Projected distribution of stars at $T=9.82$ Myr. 
 The orbital plane of the star cluster has an inclination of
 $i=127^{\circ}$ with respect to the plane of the sky, with a half-line
 of ascending nodes at $\Omega = 99^{\circ}$ east of north. These
 values are used to mimic the result of
 \citet{2006ApJ...643.1011P}.  Arrows show the proper motion of stars. 
 Red and white crosses show the positions of the
 IMBH and SMBH, respectively. Dotted lines show the orbit of stars which
 reached inner 0.05 pc from the SMBH (projected distance).}
 \label{fig:snapshot}
\end{figure}

\begin{figure}[htbp]
\epsscale{0.9}
\plotone{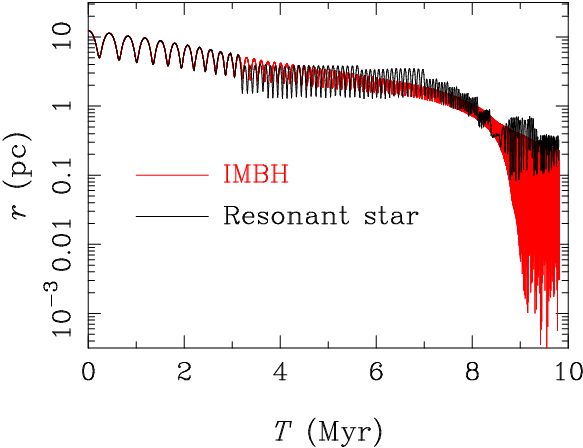} 
 \caption{Orbital evolutions of the IMBH (red) and a resonant star (black).}
 \label{fig:IMBH_orbit}
\end{figure}

Figure \ref{fig:snapshot} shows the projected distribution of stars 
at the end of the run (9.82Myr). We plotted only the stars which were originally the
members of the star cluster. The orbital plane of the star 
cluster has an inclination of
$i=127^{\circ}$ with respect to the plane of the sky and with a half-line
of ascending nodes at $\Omega = 99^{\circ}$ east of north. These
values follow the result of \citet{2006ApJ...643.1011P}.
The arrows show the proper motions of the stars. Red arrows are clockwise
and green ones are counterclockwise. Dotted curves show the orbits of stars
which reached the projected distance less than 0.05 pc from the central 
SMBH. The distribution of stars in this snapshot is very similar to those of S-stars, 
but the scale is larger in the simulation.
As we stated above, these stars are carried by the 1:1 mean motion resonance
with the IMBH and released from the resonance when the orbit of the IMBH becomes 
highly eccentric (see figure \ref{fig:IMBH_orbit}). Thus, the radial distribution of 
the carried stars strongly depends on the radius at which the IMBH orbit becomes 
eccentric. \citet{2006MNRAS.372..174B}, \citet{2007ApJ...656..879M},
and Iwasawa et al. (preprint) showed that the orbit of IMBHs becomes highly
eccentric after their orbital decay slowed down because of the depletion of stars.
The mechanism is as follows (Iwasawa et al, preprint):
When the orbit of the IMBH is exactly circular, the orbits of
stars with the semimajor axis larger than that of the IMBH would
follow the Kozai cycle, which is a large-amplitude coupled oscillation
of eccentricity and inclination. In this case, the $z$-component of the
angular momentum, $J_z$, is conserved, since the time-averaged
potential of the SMBH and the IMBH is axisymmetric. If the orbit of the IMBH is not
exactly circular, for some stars $J_z$ is no longer conserved. Thus,
they can exchange the angular momentum with the IMBH. Moreover, they are
likely to be ejected by the IMBH when they are in the prograde orbits
with the peri-center distance comparable to the semi-major axis of the
IMBH.  On average, these stars carry away the orbital angular momentum
of the IMBH and cause an increase in the eccentricity of the IMBH.

\begin{figure}[htbp]
\epsscale{0.9}
\plottwo{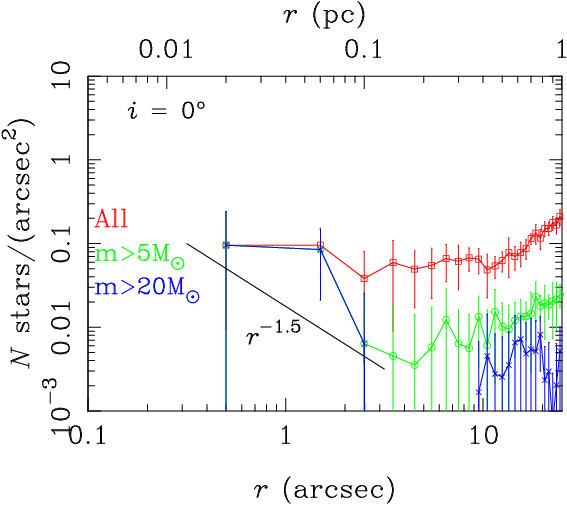}{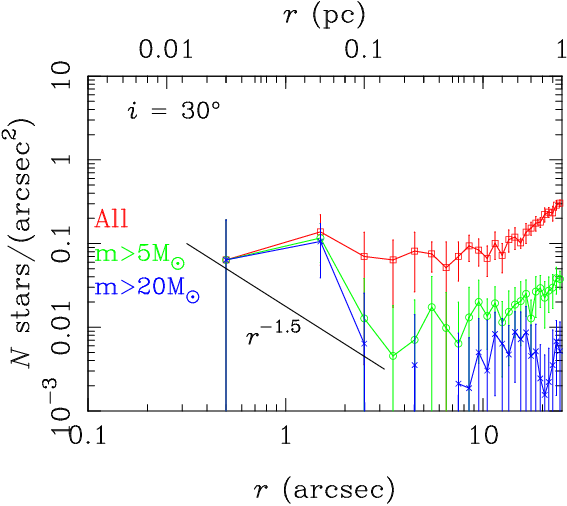}
\plottwo{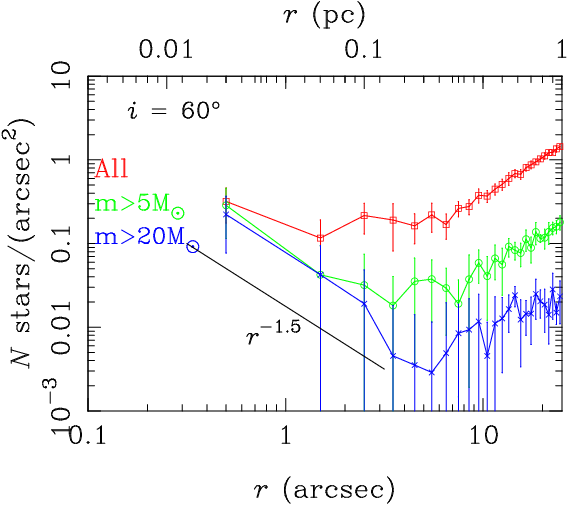}{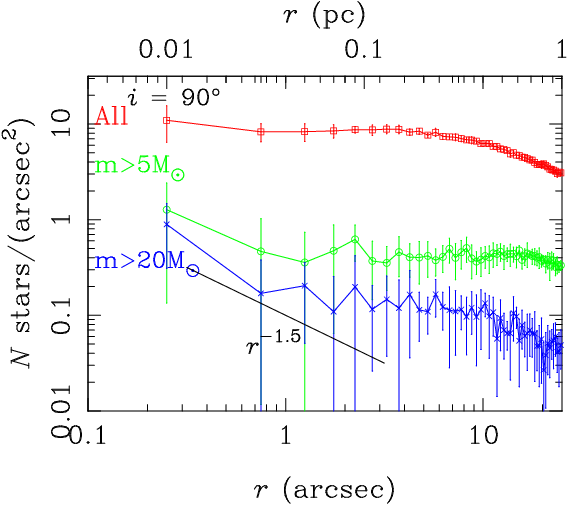}
 \caption{Surface density of young stars as a function of projected distance 
from the SMBH averaged within the last ten snapshots.
Top left, top right, bottom left, and bottom right panels show the projections with 
inclinations $i=0^{\circ}, 30^{\circ}, 60^{\circ}$, and $90^{\circ}$, respectively.
Red curves includes all stars. Green and Blue curves include stars with 
$m>5 M_{\odot}$ and $m>20 M_{\odot}$, respectively. Error bars show their standard
deviations.}
 \label{fig:sd}
\end{figure}

We investigated the surface density profile, averaged over the last
10 snapshots with a time interval of around 8000 years. Since the young stars 
distribute mainly in the orbital
plane of the star cluster, we plotted the projected surface density for
inclinations $i=0^{\circ}$, $30^{\circ}$, $60^{\circ}$, and $90^{\circ}$.
The inclination of the observed clockwise disk at 0.1--1 pc is $127^{\circ}$
\citep{2006ApJ...643.1011P}. It is similar to $i=60^{\circ}$.
Red curves in figure \ref{fig:sd} show the surface density of all
young stars. Green and blue curves show those of massive stars only with 
$m>5M_{\odot}$ and $m>20M_{\odot}$, respectively. The error bars show their 
standard deviations. While the massive stars concentrate on the GC,
the distribution of all stars is rather flat in all inclinations.
The slope of the projected surface density profile of the massive stars is around 
$-1.5$ within $10''$, although it is slightly different for the different inclinations 
and the number of particles is too small to fit in the case of small inclinations.
This result agrees well with the observed surface density of young massive 
stars \citep{2010ApJ...708..834B,2009A&A...499..483B,2009ApJ...703.1323D}.

\begin{figure}[htbp]
\epsscale{0.9}
\plotone{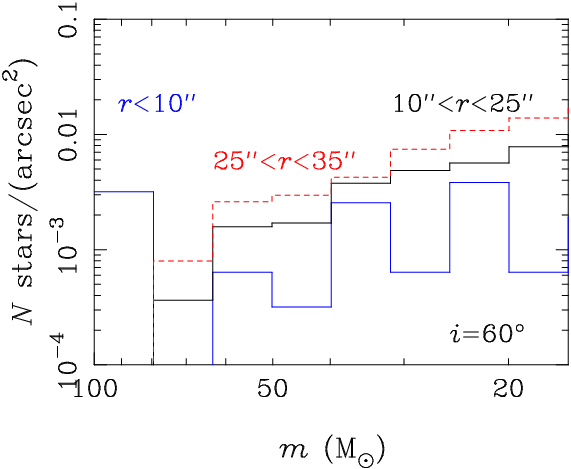} 
 \caption{Mass functions of young stars in three radial intervals, 
$r<10'', 10''<r<25''$, and $25''<r<35''$, averaged within the last ten snapshots. 
 We rotated the stellar disk with $i=60^{\circ}$ from the orbital plane of the IMBH 
 for the comparison with the observation \citep{2010ApJ...708..834B}.}
 \label{fig:mf}
\end{figure}

We also investigated the mass functions of young stars in three intervals, 
$r<10'', 10''<r<25''$, and $25''<r<35''$, averaged within the last 10 snapshots. 
Figure \ref{fig:mf} shows the result for $i=60^{\circ}$, which is a similar angle to 
the observed young stellar disk. The mass function was extremely top-heavy in the 
inner-most region, $r<10''$, and approaches a power-law mass function in the outer 
region. These results are consistent with the observation \citep{2010ApJ...708..834B}.

As shown in \citet{2009ApJ...695.1421F}, the star cluster can carry 
massive stars efficiently to the GC because of the following mechanisms. 
Massive stars sink to the cluster center because of the mass segregation. 
When a star cluster is disrupted by the tidal force of the parent galaxy, stars in 
the outer region of the cluster become unbound
first. Therefore, massive stars tend to be carried close to the GC. After the disruption
of the star cluster, some of the massive stars are carried further into the inner 
region by the 1:1 mean motion resonance of the IMBH.

\begin{figure}[htbp]
\epsscale{0.9}
\plottwo{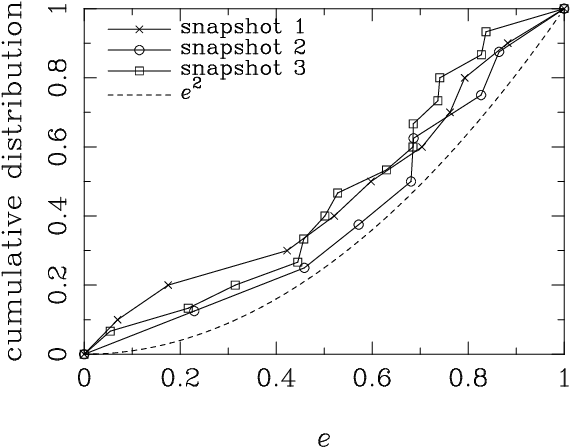}{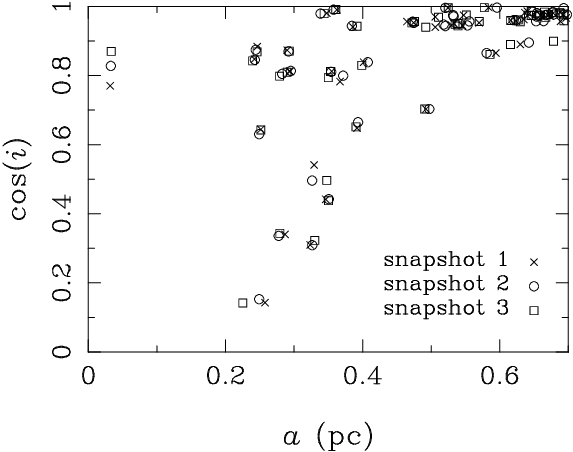} 
 \caption{Distribution of the orbits of young stars around the GC. Left: cumulative 
distribution for the eccentricities within 0.3 pc for the last three snapshots. Dashed
curve shows a thermal distribution, $n(e)=2e$. 
Right: distribution of the inclinations of young stars in the last three snapshots.}
 \label{fig:pdf_e}
\end{figure}

Figure \ref{fig:pdf_e} shows the probability distribution function
of young stars within 0.3 pc from the GC (left panel of figure \ref{fig:pdf_e})
and the distribution of inclinations (right panel of figure \ref{fig:pdf_e}). 
The distribution of the eccentricity shows a thermal distribution, $n(e)=2e$
\citep{1975MNRAS.173..729H}. This result agrees with the simulation performed 
by \citet{2009ApJ...693L..35M}.
The inclination is randomized near the GC, but no star has an inclination larger 
than 90$^\circ$ in this snapshot. However, the inclination of one resonant 
star exceeded 90$^\circ$. Several stars escaping the 
cluster by slingshot also had retrograde orbits and have the pericenter distance 
less than 0.1 pc.

\section{Summary and Discussion}

We performed a fully self-consistent $N$-body simulation of a star
cluster near the GC. The star cluster migrated to the
GC owing to the dynamical friction and was disrupted by the
tidal force. The stars which are initially the members of the star cluster 
formed a disk structure. Before the disruption, an IMBH formed in the
cluster via runaway collisions of stars. After the disruption of the
cluster, the IMBH continued to sink to the GC and stars which were caught in the 
1:1 mean motion resonance of the IMBH also sank to the GC. Near the GC, 
the spiral-in of the IMBH slowed because of the depletion of field stars, and the 
orbit of the IMBH became highly eccentric. At this stage, the stars were kicked 
out from the resonance and the orbits were efficiently randomized by the 
non-axisymmetric perturbing potential of the IMBH.

We investigated the distributions of the stars carried to the GC by the star cluster 
and the IMBH. We found that they agree well with the observed ones.
The surface density within $10''$ had a slope of $-1.5$. 
The mass function of the young stars was extremely top-heavy in the inner-most region. 
The eccentricities and inclinations of the young stars carried near the central 
SMBH by the resonance were a thermal and isotropic distribution, while young stars 
in the outer region were distributed in a disk. These distributions agree with that of
S-stars and a young stellar disk.
Thus, the distributions of ``debris stars'' of the sinking star cluster agree well 
with the observations. The star cluster scenario with an IMBH can explain the origin 
of both a young stellar disk and S-stars.

Here, we discuss possible conditions for the formation of S-stars. 
In our simulation, the distances of young stars carried by the resonance 
were larger than those of S-stars. However, how deeply the IMBH can carry 
stars to the GC depends on the distance where the orbit of the 
IMBH becomes eccentric. It occurs when the mass of the IMBH is 
comparable to the enclosed mass of the field stars 
\citep{2007ApJ...656..879M} and this distance strongly
depends on the density distribution of the GC and the mass of the IMBH. 
First, we discuss the density distribution. The enclosed mass of the field 
stars in our model is only $\sim10$\% of that estimated by observed visible
stars \citep{2007A&A...469..125S} at 0.15 pc, where the orbit of the IMBH became 
eccentric in our simulation. If the density profile is a
broken power-law fitted by observed visible stars, the enclosed mass 
is $\sim 1000 M_{\sun}$ even at 0.01 pc 
\citep{2007A&A...469..125S,2003ApJ...594..812G}.
We also estimate the case of a power-law with $-7/4$, which is theoretically 
expected \citep{1976ApJ...209..214B}. In this case, the enclosed mass is 
8000$M_{\sun}$ at 0.01 pc (2000 AU) and 1000 $M_{\sun}$ at $2\times10^{-3}$ 
pc (400 AU). These values are smaller than the
upper limit of the enclosed mass from the observation of S2, 3--4
$\times10^5M_{\odot}$ at 0.01 pc \citep{2008ApJ...689.1044G}.
Recent observations show a flat density profile of old stars
\citep{2009ApJ...703.1323D,2009A&A...499..483B,2010ApJ...708..834B}. 
However, it is unlikely that this distribution reflects the true mass 
distribution. There are certainly dark masses composed of stellar mass black 
holes, neutron stars, and white and brown dwarfs. On the other hand,
the timescale of collisions between main-sequence stars is pretty small, 
around 0.1--1 Gyr for inner 1 pc, if a 
stellar cusp developed through thermal relaxation 
\citep{1983ApJ...268..565D}. Therefore, the lack of old stars is 
probably the result of collisional disruptions. Compact objects are not
disrupted by collisions, and brown dwarfs are also less likely to be 
disrupted by collisions due to their high density. Thus, it seems 
natural to assume that the innermost region of the GC is
dominated by dark mass.

Next, we discuss the possible range of the IMBH mass. 
In our simulation, the IMBH is more massive than the observational upper 
limit, $\sim 10^4 M_{\odot}$ \citep{2004ApJ...616..872R}, but it is 
possible to form smaller IMBHs from different initial conditions or with 
higher mass 
loss rates \citep{2009ApJ...695.1421F}. Even if we assume an extreme mass
loss rate, it is possible to form a star with a few thousand solar masses through
runaway collisions because the star cluster near the GC is very compact.
For the lower limit, $1500 M_{\odot}$ 
is sufficient for the randomization of stars \citep{2009ApJ...693L..35M}. 
Less massive IMBHs can carry stars closer to the GC, but it takes a longer time 
to migrate due to the dynamical friction. 
Star clusters migrate to around 1pc and are disrupted there. Its typical 
timescale is 2--10 Myr for a star cluster at 5--10 pc from the GC with around 
$10^5 M_{\sun}$\citep{2009ApJ...695.1421F}. 
After the disruption, IMBHs migrate due to the dynamical friction. We 
estimated the timescale for IMBHs migrating from 
1 pc to $10^{-3}$ pc using the equation derived from \citet{2007ApJ...656..879M}. 
We assumed that the Bahcall--Wolf cusp and obtained 1 Myr for the IMBH with 
16000$M_{\odot}$. This result is consistent with our simulation. We also tried 
broken power laws \citep{2003ApJ...594..812G} and obtained similar results. 
The timescale is inversely proportional to the IMBH mass. 
Thus, IMBHs with a few thousand solar masses are capable for this scenario.

\acknowledgments
We thank Keigo Nitadori for NINJA and Phantom-GRAPE with high
accuracy and Tomoaki Ishiyama and Kuniaki Koike for helpful discussion on
our new code.
M.F. is financially supported by Research Fellowships of JSPS for Young
Scientist. 
Numerical computations were carried out on Cray XT4 at the Center for
Computational Astrophysics, CfCA, of National Astronomical Observatory
of Japan.

\onecolumn
\twocolumn

\onecolumn
\twocolumn

\end{document}